#########################################################################

\documentstyle[12pt]{article}
\begin{document}
\def\be{\begin{equation}}
\def\ee{\end{equation}}
\def\ds{\displaystyle}

\font\tencmsa=msam10
\font\sevencmsa=msam7
\font\fivecmsa=msam5
\newfam\sa
\textfont\sa=\tencmsa
\scriptfont\sa=\sevencmsa
\scriptscriptfont\sa=\fivecmsa
\def\hexnumber@#1{\ifcase#1 0\or1\or2\or3\or4\or5\or6\or7\or8\or9\or
 A\or B\or C\or D\or E\or F\fi}
\edef\san@{\hexnumber@\sa}
\mathchardef\lsim="1\san@2E
\mathchardef\lleq="1\san@36
\mathchardef\ggeq="1\san@3E
\mathchardef\gsim="1\san@26

\begin{center}
{\Large\bf Non-perturbative beta-function in $SU(2)$ lattice
gauge fields thermodynamics}\\

\vspace*{1cm}

{\bf O.~Mogilevsky\footnote{email: mogil@ap3.bitp.kiev.ua}, }\\

\vspace*{0.5cm}
{\large \it
N.N.Bogolyubov Institute for Theoretical Physics, National Academy
of Sciences of Ukraine, 252143 Kiev, Ukraine}\\
\vspace*{0.3cm}

\end{center}
\vspace{.5cm}
\begin{abstract}
The new method of nonperturbative calculation of the beta function in
the lattice gauge theory is proposed. The method is based on the
finite size scaling hypothesis.
\end{abstract}
\vspace*{0.3cm}

Monte Carlo simulations of SU(2) lattice gauge theory have shown that
there are rather large deviations from asymptotic scaling behaviour in
the range of coupling constants accessible with today's computing power.
This raised the question of whether we are able to see continuum
physics in these simulations. The answer requires the knowledge of the
$\beta$-function away from the asymptotic regime where it is dominated
by the two leading terms in its perturbative expansion ($g$-coupling
constant)
\be
\beta_f(g)=-b_0g^3-b_1g^5-\ldots,
\ee
where ${\ds b_0={11\over24\pi^2}}$, ${\ds b_1={17\over96\pi^4}}$.

\noindent
During the last years a large effort has been put into the numerical
determination of the $\beta$-function for the lattice gauge theories
by measuring the deviation from the two-loop $\beta$-function (1)
[1,2].

We propose a new method of determination of the nonperturbative
$\beta$-function, which is based on the finite size scaling 
hypothesis.

We consider SU(2) gauge theory at finite temperature on
$N_\sigma^3\times N_\tau$ lattices with the standard Wilson action
\be
S(U)={4\over g^2}\sum_{p}\biggl(1-{1\over2}{\rm Tr}\, U_p\biggr),
\ee
where $U_p$ is the product of link operators around the plaquette. The
number of lattice points in the space (time) direction
$N_\sigma(N_\tau)$ and the lattice spacing $a$ fix the volume and
temperature
\be
V=(N_\sigma a)^3,\qquad T=1/(N_\tau a).
\ee
The $\beta$-function is defined by the expression
\be
\beta_f(g)=-a{dg(a)\over da}.
\ee
Then for lattice spacing $a$  one can obtain
\be
a={1\over\Lambda_L}\exp\biggl(-\int {dg\over\beta_f(g)}\biggr)
\ee
where $\Lambda_L$ is the renormalization group invariant parameter.
Since in Monte Carlo simulations the thermodynamic functions are
calculated in the units of  lattice spacing $a$, formula (5)
determines the temperature dependence of these functions according (3).

In the asymptotically free (AF) regime (1) we obtain the well known formula
\be
a\Lambda_L = R(g^2) = 
\exp\biggl\{-{b_1\over2b_0^2}\ln(b_0g^2)-{1\over2b_0g^2}\biggr\} \ ,
\ee
which is valid in the region $g^2<1$.

\noindent
On the other hand, SU(2) gauge system at finite $N_{\tau}$ and 
undergoes the deconfinement
phase transition at $g^2\gsim 1$. The new nonperturbative method for
the calculation of $\beta$-function is needed, which is not connected
with the expansion (1).

Our approach is based on the two points: i) translation into a more
conventional statistical mechanical definition of $\beta$-function and
ii) the finite size scaling theory and phenomenological renormalization.
As in the standard spin systems, let us make the infinitesimal
transformation of the lattice spacing $a\to a'=ba=(1+\Delta b)a$. Then
\be
-a{dg\over da}=-\lim_{b\to 1}\biggl(a\cdot{g(ba)-g(a)\over
ba-a}\biggr)=-\lim_{b\to 1}{\Delta g\over \Delta b}=
-\lim_{b\to 1}{dg\over db}.
\ee
We obtain the new definition of $\beta$-function for SU(2) lattice
gauge system
\be
\beta_f(g)=
-\lim_{b\to 1}{dg\over db}.
\ee
It has been shown in the finite size scaling theory that on the finite
lattice $N_\sigma^3\times N_\tau$ ($N_\tau$ fixed) the order parameter
$\langle { L}\rangle$,
the susceptibility $\chi$ and the
correlation  length $\xi$ can be expressed in the following form
(see, for example [3])
\be
O(g^{-2},N_\sigma)=N_\sigma^{{\omega\over\nu}}Q_0(g^{-2},N_\sigma).
\ee
Here $O$ represents
$\langle { L}\rangle$,
$\chi$ and $\xi$, $\omega=-\beta,\gamma,\nu$ is the corresponding
critical index. Scaling function $Q$ has some special dependence on
$g^{-2}$ and $N_\sigma$, but this is out of our consideration.

\noindent
For example
\be
\left\{
\begin{array}{ccl}
\langle{ L}\rangle&=&N_\sigma^{-{\beta\over\nu}}Q_{\langle{ L}\rangle}
(g^{-2},N_\sigma),\\
\xi&=&N_\sigma Q_\xi
(g^{-2},N_\sigma).
\end{array}\right.
\ee
The existence of the scaling function $Q$ allows to develop a
procedure to renormalize the coupling constant $g^{-2}$ by the use of
two different lattice sizes $N_\sigma$ and $N'_\sigma$. Let us fix the
physical size ${ L}=N_\sigma a$ and make a scale transformation
\be
\left\{
\begin{array}{l}
a\to a'=ba,\\ N_\sigma\to N'_\sigma=N_\sigma/b.
\end{array}\right.
\ee
Then the phenomenological renormalization is defined by the equation
\be
Q(g^{-2},N_\sigma)=Q\bigl((g')^{-2},N_\sigma/b\bigr)
\ee
It expresses that the scaling function $Q$ remains unchanged if the
lattice size is rescaled by a factor $b$ and the inverse coupling
$g^{-2}$ is shifted to $(g')^{-2}$ simultaneously. Taking the
derivative with respect to the scale parameter $b$ of the both sides of
(12) and using (8) it is easy to obtain the expression
\be
a{dg^{-2}\over da}={\partial\ln Q/\partial\ln N_\sigma\over \partial\ln Q/
\partial g^{-2}}.
\ee
The approximation of the derivative with respect to $N_\sigma$ by the
finite difference yields the final formula for the $\beta$-function
\be
a{dg^{-2}\over da}={\ln\biggl[{\ds{Q(N'_\sigma)\over
Q(N_\sigma)}}\biggr]\biggm/\ln\biggl({\ds{N'_\sigma\over
N_\sigma}}\biggr)\over\biggl[{\ds{dQ(N_\sigma)\over
dg^{-2}}}{\ds{dQ(N'_\sigma)\over
dg^{-2}}}\biggm/Q(N_\sigma)Q(N'_\sigma)\biggr]^{1/2}}\,.
\ee

So far we have considered the scaling function $Q$. 
It is very interesting to apply the same analysis to the
correlation length $\xi$. Using (10) we obtain instead of (12)
\be
{\xi(g,N_\sigma)\over N_\sigma}
=
{\xi(g',N'_\sigma)\over N'_\sigma}
\ee
In the case of the large enough lattice size $(N_\sigma\to\infty)$ the
dependence of the correlation length $\xi$ on $N_\sigma$ become
negligible and (15) yields
\be
\xi(g)=b\xi(g').
\ee
This is the renormalization group equation for the bulk system, which
is wellknown in the standard spin theory. From this by the
manipulations mentioned above one can obtain the expression for the
$\beta$-function
\be
\beta_f(g)=\biggl[{d\ln\xi(g)\over dg}\biggr]^{-1}.
\ee
Substituting (17) into (5) we have an  extremely simple formula for
the lattice spacing \be
a\Lambda_L={1\over\xi(g)}.  \ee
The first attempt to calculate the correlation length $\xi(g)$ has
been made in [4], but with rather pure statistics and on rather small
lattices $(N_\sigma=18$, $N_\tau=3,4,5)$. We can see that
the new method of calculation of the
nonperturbative $\beta$-function for SU(2) lattice gauge
thermodynamics only needs the correct calculation of the
correlation length $\xi$ in the wide coupling constant interval.

On the other hand, the best studied quantities in MC lattice calculations 
of $SU(N)$ gauge theories are the string tension $\sqrt{\sigma}$  and the
deconfinement transition temperature $T_c$. The MC calculations give
the values of the critical coupling $\beta_c^{MC}$ of the
deconfinement transition and dimensionless string tension
$(\sqrt{\sigma})_{MC}$. The values $\beta_c^{MC}$ were found for the
finite lattices and the extrapolation to spatially infinite volume
(``thermodynamical limit'') $N_\sigma\to\infty$ has been done (see
Ref. [5] and references therein). For the $SU(2)$ gauge theory the MC
values of the critical couplings $\beta_c^{MC}$  are presented in
Table 1 for different $N_\tau$. 
One observes a rather strong dependence of
$T_c/\Lambda_L^{AF}$ on $N_\tau$. 
This means that the perturbative AF relation (6) does not
work even on the largest available lattices. This fact is  known as an
absence of the asymptotic scaling.

It has been proposed in Ref. [2] that a deviation from the asymptotic
scaling can be described by a universal non-perturbative (NP) beta
function, i.e. $\beta_f(g)$ is the same for all lattice observables
and it does not depend on the lattice size if $N_\sigma$, $N_\tau$
are not too small. The following ansatz was suggested [2]:
\be
a\Lambda_L^{NP}=\lambda(g^2)R(g^2),
\ee
where $R(g^2)$ is given by Eq. (6) and $\lambda(g^2)$ is thought to
describe a deviation from the perturbative behaviour. The equation (6)
has been expected at $g\to0$ so that an additional constraint,
$\lambda(0)=1$, has been assumed. The values of
$T_c/\Lambda_L^{NP}$
can be calculated then as
\be
T_c/\Lambda_L^{NP}={1\over N_\tau\lambda(g_c^2)R(g_c^2)}
\ee
A simple formula for the function $\lambda(g^2)$ was suggested [2]:
\be
\lambda(g^2)=\exp\biggl({c_3g^6\over2b_0^2}\biggr)
\ee
Parameter $c_3$ in (21) and a new one,
$T_c^*/\Lambda_L^{NP}={\rm const}$, were considered as free parameters
and determined from fitting the MC values of
$T_c/\Lambda_L^{NP}$
at different $N_\tau$  to the constant value
$T^*_c/\Lambda_L^{NP}$. This procedure gives:
\be
T^*_c/\Lambda_L^{NP}=21.45(14), \quad c_3=5.529(63)\cdot10^{-4}
\ee

In comparison to
$T_c/\Lambda_L^{AF}$
the much weaker $N_\tau$ dependent values of
$T_c/\Lambda_L^{NP}$ have been obtained, which become now close to the
constant value
$T^*_c/\Lambda_L^{NP}$ (22).

In spite of the phenomenological success of the above procedure of
[2] the crusial question regarding the existence of the universal NP
beta function, with does not depend on the lattice size, is not solved
and remains just a postulate. To answer this question we reanalyze the
same MC data using the different strategy as in Ref. [2]. A principal
difference of our analysis is that we do not assume the existence of
the universal beta function and take into account the finite size
effects of the lattice.

Usually finite size scaling (FSS) in the vicinity of a
finite-temperature phase transition is discussed for lattice SU(N)
gauge models, without trying to make contact with the continuum limit,
i.e. the scaling properties are studied on lattices of $N_\tau\times
N_\sigma^3$ with fixed $N_\tau$  and varying $N_\sigma$, and the model
is viewed as a 3-dimensional spin system. In the continuum limit the
FSS properties of these non-abelian models should, of course, be
discussed in terms of the physical volume $V=L^3$ and the temperature
$T$ in the vicinity of the deconfinement transition temperature $T_c$.
We will study here how the scaling behaviour of the continuum theory
emerges from the lattice free energy on arbitrary lattices, i.e. when
varying  $N_\tau$  and $N_\sigma$.

On a lattice $N_\tau\times N_\sigma^3$ the length scale $L$ and the
temperature $T$ are given in units of the lattice spacing $a$,
therefore it is advantageous to replace the length scale $L$ by the
dimensionless combination
\be
LT={N_\sigma\over N_\tau}
\ee
Using this ratio the singular part of the free energy density is
described by a universal finite-size scaling function [3,5]
\be
f_s(t,h,N_\sigma,N_\tau)=
\biggl({N_\sigma\over N_\tau}\biggr)
^{-3}Q_{f_s}\left(t
\biggl({N_\sigma\over N_\tau}\biggr)
^{1/\nu},\,h
\biggl({N_\sigma\over N_\tau}\biggr)
^{\beta+\gamma\over\nu}\right),
\ee
where $\beta$, $\gamma$, $\nu$ are the critical indexes of the theory,
the scaling function  $Q_{f_s}$  depends on the reduced temperature
$t=(T-T_c)/T_c$ and the external field strength $h$.

Next we consider $y=N_\sigma/N_\tau$ fixed, varying $N_\sigma$ and
therefore $N_\tau$ accordingly as is needed to reach the continuum
limit. Rescaling $N_\sigma$ and $N_\tau$ by a factor $b$ leads to a
phenomenological renormalization $g'(g,b,y)$ by the following identity
for a scaling function $Q$
\be
Q\left(t(g,N_\tau)\cdot\biggl({N_\sigma\over
N_\tau}\biggr)^{1/\nu}\right)=
Q\left(t(g',N_\tau/b)\cdot\biggl({bN_\sigma\over
bN_\tau}\biggr)^{1/\nu}\right).
\ee
It follows from (25)
\be
t(g,N_\tau)=t(g',N_\tau/b).
\ee
In general the reduced temperature $t=(T-T_c)/T_c$ is a complicated
function of the coupling $\beta=2N/g^2$, which in the visinity of the
critical temperature $T_c$ can be approximated by [5]
\be
t=(\beta-\beta_c){1\over4Nb_0}\biggl[1-{2Nb_1\over
b_0}\beta_c^{-1}\biggr]
\ee
This approximation reproduces the correct reduced temperature in the
continuum limit, which is easy verified by using (4). Taking the
derivative with respect to the scale parameter $b$ of the both sides
of (26) and using (8) and (27) it is easy to obtain the expression
for the beta function:
\be
\beta_f(g)=-B_0(N_\tau)g^3-B_1(N_\tau)g^5,
\ee
where
\be
\left\{\begin{array}{l}
B_0(N_\tau)=\ds{1\over 4N}\biggl(1-\ds{2Nb_1\over
b_0\beta_c}\biggr)\ds{d\beta_c\over d\ln N_\tau} \\
B_1(N_\tau)=B_0(N_\tau)\ds{b_1\over b_0}\end{array}\right.
\ee
Then the equation (4) leads to
\be
a\Lambda_L=\exp\biggl(-{1\over2B_0g^2}\biggr)(B_0g^2)^{-B_1/2B_0^2}.
\ee
Using (3) and (30) one can obtain the critical temperature $T_c$. The
only problem is the calculation of the derivative $d\beta_c/d\ln
N_\tau$ in expression (29). For the $SU(2)$ gauge theory the
calculation has been made by fitting the MC data for the critical
couplings $\beta_c^{MC}$ with a Spline interpolation and  numerical
differentiation of this curve, The result of the calculation is
represent in Table 1. In comparison to $T_c/\Lambda_L^{AF}$
the much weaker dependence on $N_\tau$ of the critical temperature
$T_c/\Lambda_L$  is observed.


\vspace{0.5cm} 
\begin{center}
{\bf Acknowledgements} 
\end{center}

Author would like to thank O.~Borisenko and M.~Gorenstein for
interesting discussions.

\renewcommand{\arraystretch}{1.3}
\begin{table}[htb]
{\baselineskip=11pt
\caption[]{
 MC data for $\beta_c$ are taken from [5]. The values of
$T_c/\Lambda_L^{AF}$
are calculated from (6). Our results for
$T_c/\Lambda_L$ are obtained from (30).
\baselineskip=10pt
}}
\begin{center}
\begin{tabular}{r|c|c|c|c}
\hline\hline
&&&&\\[-.4cm]
$N_\tau$&$\ds{\beta_c={4\over
g_c^2}}$&$\ds{T_c/\Lambda_L^{AF}}$&$\ds{d\beta_c\over
dN_\tau}$&$T_c/\Lambda_L$
\\
&&&&\\[-.4cm]
\hline
2&1.880&29.7&---&---\\
3&2.177&41.4&0.158&25.22\\
4&2.299&42.1&0.086&25.46\\
5&2.373&40.6&0.063&25.38\\
6&2.427&38.7&0.045&24.13\\
8&2.512&36.0&0.040&24.24\\
16&2.739&32.0&0.017&---\\
\hline\hline
\end{tabular}
\end{center}
\end{table}

\end{document}